\documentclass[aps,prd,twocolumn,showpacs,preprintnumbers,amsmath,amssymb]{revtex4}

\usepackage{bbm}
\usepackage{graphicx}
\usepackage{amsmath}
\usepackage{amsfonts}
\begin{document}
\title{Light Pseudoscalar Meson and Heavy Meson Scattering Lengths}

\author{Yan-Rui Liu }
\email{yrliu@ihep.ac.cn} \affiliation{Institute of High Energy
Physics, P.O. Box 918-4, Beijing 100049, China}
\author{Xiang Liu }
\affiliation{School of Physical Science and Technology, Lanzhou
University, Lanzhou 730000, China and \\
Centro de F\'{i}sica Computacional, Departamento de F\'{i}sica,
Universidade de Coimbra, P-3004-516 Coimbra, Portugal}
\author{Shi-Lin Zhu }
\email{zhusl@phy.pku.edu.cn}
\affiliation{Department of Physics,
Peking University, Beijing 100871, China}

\begin{abstract}
We have performed a systematical calculation of the pion ($\pi, K,
\eta$) and heavy pseudoscalar meson S-wave scattering lengths up to
${\cal O}(p^3)$ in the chiral perturbation theory in the heavy quark
symmetry limit. With the three scattering lengths from the lattice
simulations as input we estimate the unknown low-energy constants.
Then we predict all the other unmeasured scattering lengths. The
analytical expressions and predictions may be helpful to future
investigations. Especially we note that the DK scattering length is
positive. Therefore their interaction is attractive, which helps to
lower the mass of the ``bare'' charm-strange scalar state in the
quark model through the couple-channel effect.
\end{abstract}


\maketitle

\section{Introduction}\label{sec1}

The discovery of the narrow $D_{sJ}(2317)$ has inspired heated
discussions of its structure in the past six years
\cite{ds2317-babar,ds2317-belle,ds2317-cleo}. The possible
interpretations include the chiral partners of $D_s$ \cite{Bardeen},
P-wave excited states of $D_s(D_s^*)$ \cite{Chao}, couple-channel
effects between the $c\bar s$ state and $DK$ continuum
\cite{Beverenn1}, conventional $c\bar s$ states \cite{dai0,Narison},
four-quark states
\cite{Chen,Cheng,TBarnes,2317-lutz,2317-guo,2317-oset,zzy} etc. (for
a detailed review see Ref. \cite{review}).

Considering the large contribution of the S-wave $DK$ continuum,
the mass of $D_{sJ}^{*}(2317)$ agrees well with the experimental
value, which is indicated by the coupled-channel effect
\cite{Beverenn1} and  the QCD sum rule (QSR) approach \cite{Dai}.
Therefore, $D_{sJ}^*(2317)$ is very probably a conventional
$c\bar{s}$ state with $J^P=0^+$. Since $D_{sJ}(2317)$ strongly
couples to the $DK$ channel, the study of the $DK$ interaction is
very interesting.

The scattering length is an important observable, which encodes
the information of the underlying interaction. For example, a
positive scattering length suggests there exists attraction in
this channel. In this work we investigate the scattering lengths
in the pion-heavy meson channels. Here the pion denotes $\pi$, $K$
or $\eta$ while the heavy mesons are the pseudoscalar charmed or
bottom mesons.

There are several works on these scattering lengths in the
literature. The S-wave $DK$ scattering length was predicted to be
$5\pm1$ GeV$^{-1}$ in a unitarization model \cite{beve}. The
S-wave scattering lengths $a_{B\pi}^{(1/2)}=0.26(26)m_\pi^{-1}$,
$a_{D\pi}^{(1/2)}=0.29(4)m_\pi^{-1}$ were extracted from a lattice
calculations of the scalar form factors in the semileptonic decays
\cite{nieves}. Recently, a lattice study on the scattering lengths
of the light hadrons with the charmed mesons and charmonia were
performed in full QCD \cite{latDK}. Unfortunately their pion
masses are still quite large. The scattering lengths of heavy
mesons and Goldstone bosons were discussed to the next leading
order in chiral perturbation theory recently \cite{guo}. In this
paper, we will go to the next-next-leading order of the chiral
expansion and consider the important loop corrections to the
elastic pion-heavy meson scattering at threshold in the heavy
quark limit.

It is known that the chiral perturbation theory works well for the
light pseudoscalar meson systems. For the meson-baryon
interactions, the heavy baryon chiral perturbation theory
(HB$\chi$PT) was proposed so that a systematic power counting rule
exists. The scattering lengths in $\pi N$, $KN$ and other channels
have been investigated to high orders within this framework
\cite{piN,kaonN,mbsl,dmbsl}. It was observed that the chiral
expansion in the SU(3) case converges well only in few channels.

For the heavy mesons, heavy quark symmetry \cite{hqs} imposes that
the ground states $D$, $D_s$, $D^*$ and $D_s^*$ belong to the same
doublet. Similar to the meson-baryon case, the heavy meson chiral
perturbation theory (HM$\chi$PT) is a useful tool to study their
interactions \cite{hmeft,hmlrev}. In the heavy quark symmetry
limit, the recoil order corrections are neglected. One may use a
power counting rule similar to that in HB$\chi$PT to include the
chiral corrections order by order. We will adopt the HM$\chi$PT
formalism to calculate the scatting lengths up to ${\cal O}(p^3)$.
Such a study may be helpful to test the convergence of HM$\chi$PT
in the scattering processes.

In our formalism, the S-wave scattering length is defined through
\begin{equation}
T_{th}=8\pi(1+\frac{m}{M})a
\end{equation}
where $T_{th}$ is the threshold T-matrix element and $m$ ($M$)
denotes the light (heavy) meson mass.

This paper is organized as follows. In Sec. \ref{sec2}, we present
the chiral Lagrangians for the calculation at threshold. In Sec.
\ref{sec3}, we give the threshold T-matrices for the S-wave
elastic scattering of the Goldstone bosons and charmed mesons up
to the third order. According to the heavy quark symmetry, these
T-matrices are also applicable to the pion-bottom meson case. In
Sec. \ref{sec4}, we estimate the values of the scattering lengths.
The final section is the discussion.

\section{Lagrangians}\label{sec2}

The lowest chiral Lagrangian for the light pseudoscalar mesons
reads
\begin{equation}
{\cal L}^{(2)}_{\phi\phi}=f^2 {\rm tr}(u_\mu
u^\mu+\frac{\chi_+}{4}),
\end{equation}
where $f\approx92.4$ MeV is the pion decay constant in the chiral
limit, $u_\mu$ is the axial vector field with the definition
\begin{equation}
u_\mu={i\over 2} \{\xi^\dagger, \partial_\mu \xi\},\quad \xi =
\exp(i \phi/2f),
\end{equation}
\begin{eqnarray}
\phi=\sqrt2\left(
\begin{array}{ccc}
\frac{\pi^0}{\sqrt2}+\frac{\eta}{\sqrt6}&\pi^+&K^+\\
\pi^-&-\frac{\pi^0}{\sqrt2}+\frac{\eta}{\sqrt6}&K^0\\
K^-&\bar{K}^0&-\frac{2}{\sqrt6}\eta
\end{array}\right),
\end{eqnarray}
and
\begin{equation}
\chi_\pm = \xi^\dagger\chi\xi^\dagger\pm\xi\chi\xi,\quad
\chi=\mathrm{diag}(m_\pi^2,\, m_\pi^2,\, 2m_K^2-m_\pi^2).
\end{equation}

The doublet of ground state heavy mesons reads
\begin{eqnarray}
&H=\frac{1+v\!\!\!/}{2}\left(P^*_\mu\gamma^\mu+iP\gamma_5\right),&\nonumber\\
&\bar{H}=\gamma^0H^\dagger\gamma^0=\left(P^{*\dag}_\mu\gamma^\mu+iP^\dag\gamma_5\right)\frac{1+v\!\!\!/}{2},&\nonumber\\
&P=(D^0, D^+, D_s^+), \quad P^*_\mu=(D^{0*}, D^{+*}, D_s^{+*})_\mu,&
\end{eqnarray}
or
\begin{eqnarray}
&P=(B^-, \bar{B}^0, \bar{B}_s^0), \quad P^*_\mu=(B^{-*},
\bar{B}^{0*}, \bar{B}_s^{0*})_\mu,&
\end{eqnarray}
where $v_\mu=(1,0,0,0)$ is the heavy meson velocity. The leading
order chiral Lagrangian for the heavy mesons in the heavy quark
symmetry limit is
\begin{eqnarray}
{\cal L}^{(1)}_{H\phi}=-\langle (i\partial_0 H)\bar{H} \rangle+
\langle H\Gamma_0 \bar{H} \rangle+g\langle Hu_\mu\gamma^\mu \gamma_5
\bar{H}\rangle,
\end{eqnarray}
where $\Gamma_\mu = {i\over 2} [\xi^\dagger, \partial_\mu\xi]$ is
the chiral connection, $\langle...\rangle$ means the trace for
Gamma matrices and the summation over flavor indices is implicit.
The heavy fields have been redefined with a factor $\sqrt M$ and
have the mass dimension 3/2.

For the calculation at threshold, we need the following ${\cal
O}(p^2)$ Lagrangian
\begin{eqnarray}
{\cal L}^{(2)}_{H\phi}&=&c_0\langle H\bar{H}\rangle{\rm
tr}(\chi_+)-c_1\langle H\chi_+ \bar{H}
\rangle\nonumber\\
&&-c_2\langle H\bar{H}\rangle{\rm tr}(u_0u_0)-c_3\langle
Hu_0u_0\bar{H}\rangle.
\end{eqnarray}
The signs of these terms are consistent with those in Ref.
\cite{HLutz}.

Similar to the chiral Lagrangians for the meson-baryon systems
\cite{pinL,MBL}, there are many terms in the third order
Lagrangian. For simplicity, we write down only the piece relevant
to our threshold calculation:
\begin{eqnarray}
{\cal L}^{(3)}_{H\phi}=\kappa\langle H[\chi_-,u_0] \bar{H} \rangle.
\end{eqnarray}

\section{Threshold T-matrices}\label{sec3}

We omit details and present the explicit expressions for threshold
T-matrices to ${\cal O}(p^3)$ of the chiral expansion. The
numerical results will be estimated in the next section. At the
leading order, we have Weinberg-Tomozawa terms
\begin{eqnarray}
&T_{D{K}}^{(1)}=0,\quad T_{D{K}}^{(0)}=\frac{2m_K}{f_K^2},&
\nonumber \\
& T_{D\bar{K}}^{(1)}=-\frac{m_K}{f_K^2},\quad
T_{D\bar{K}}^{(0)}=\frac{m_K}{f_K^2},&\nonumber\\
&T_{D_sK}=-\frac{m_K}{f_K^2},\quad
T_{D_s\bar{K}}=\frac{m_K}{f_K^2},&\nonumber\\
&T_{D\pi}^{(3/2)}=-\frac{m_\pi}{f_\pi^2},\quad
T_{D\pi}^{(1/2)}=\frac{2m_\pi}{f_\pi^2},&\nonumber\\
&T_{D_s\pi}=T_{D\eta}=T_{D_s\eta}=0,&
\end{eqnarray}
with $K=(K^+, K^0)^T$, $\bar{K}=(\bar{K}^0,K^{-})^T$. The
superscripts of $T$-matrices denote the total isospin. Here we have
replaced $f$ with the renormalized decay constants $f_\pi$, $f_K$
and $f_\eta$. The resulting ${\cal O}(p^3)$ corrections will be
taken into account later. These expressions are consistent with the
results in Ref. \cite{guo} considering the different overall sign
for $T$.

It is convenient to define two combinations of the low energy
constants (LECs) at the next-leading order
\begin{eqnarray}
C_1&=&8c_0-4c_1+2c_2+c_3,\nonumber\\
C_0&=&8c_0+4c_1+2c_2-c_3.
\end{eqnarray}
From ${\cal L}^{(2)}_{H\phi}$, we get the following T-matrices
\begin{eqnarray}
&T_{D{K}}^{(1)}=\frac{m_K^2}{2f_K^2}(C_1+C_0),\quad
T_{D{K}}^{(0)}=\frac{m_K^2}{f_K^2}C_1,&
\nonumber \\
& T_{D\bar{K}}^{(1)}=\frac{m_K^2}{f_K^2}C_1,\quad
T_{D\bar{K}}^{(0)}=\frac{m_K^2}{f_K^2}C_0,&\nonumber\\
&T_{D_sK}=\frac{m_K^2}{f_K^2}C_1,\quad
T_{D_s\bar{K}}=\frac{m_K^2}{f_K^2}C_1,&\nonumber\\
&T_{D\pi}^{(3/2)}=\frac{m_\pi^2}{f_\pi^2}C_1,\quad
T_{D\pi}^{(1/2)}=\frac{m_\pi^2}{f_\pi^2}C_1,&\nonumber\\
&T_{D_s\pi}=\frac{m_\pi^2}{2f_\pi^2}(C_1+C_0)&\nonumber\\
&T_{D\eta}=\frac{1}{3f_\eta^2}\left[(2C_1+C_0)m_\eta^2+4c_1(m_\eta^2-m_\pi^2)\right],&\nonumber\\
&T_{D_s\eta}=\frac{1}{6f_\eta^2}\left[(7C_1-C_0)m_\eta^2-16c_1(m_\eta^2-m_\pi^2)\right].&
\end{eqnarray}
The effects due to the decay constant renormalization are beyond
the order we are considering.

\begin{figure*}
\begin{center}
\includegraphics{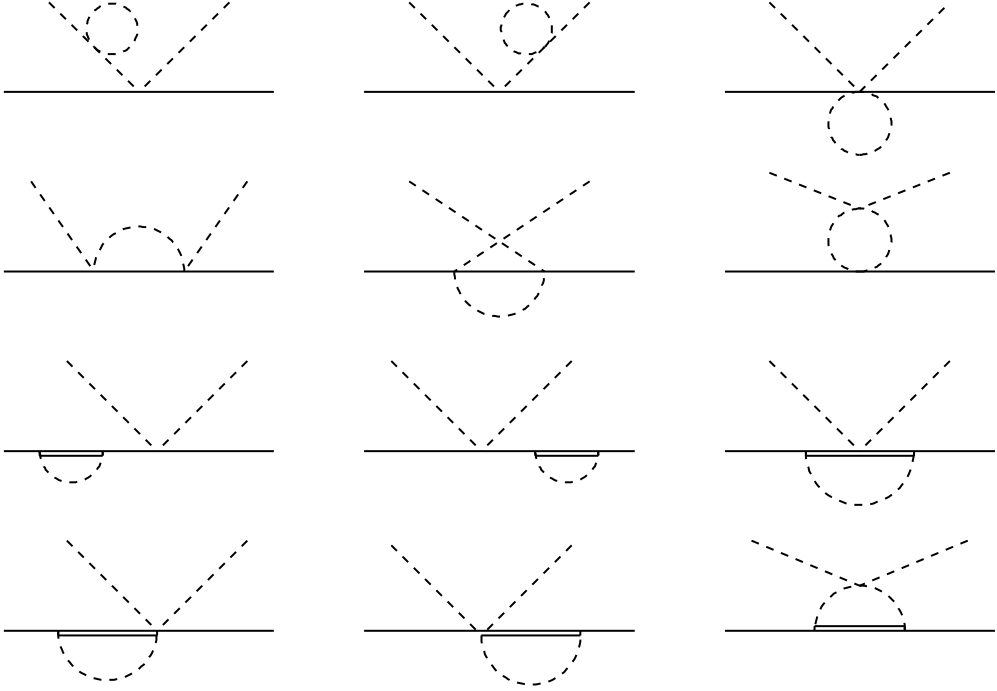}
\end{center}
\caption{Non-vanishing loop diagrams in the calculation of
meson-heavy meson scattering lengths to the third chiral order in
HM$\chi$PT. Dashed lines represent Goldstone bosons while solid
(double) lines represent pseudoscalar (vector) heavy mesons. The
fourth diagram generates imaginary parts for kaon-heavy meson and
eta-heavy meson scattering lengths.}\label{figg}
\end{figure*}

At the third order, we have the non-vanishing corrections from the
loop diagrams in Fig. \ref{figg}.

\begin{eqnarray}
T_{DK}^{(1)}&=&\frac{m_K^2}{8\pi^2f_K^4}\left\{-m_K\left(\ln\frac{m_\pi}{\lambda}
-\ln\frac{|m_K|}{\lambda}\right)\right.\nonumber\\
&&\left.+\sqrt{m_K^2-m_\pi^2}\left(i\pi-\ln\frac{m_K+\sqrt{m_K^2-m_\pi^2}}{m_\pi}\right)\right.\nonumber\\
&&\left.-\frac{1}{6}g^2\pi
\left(m_\eta+\frac{2m_\pi^2}{m_\eta+m_\pi}\right) \right\},
\end{eqnarray}
\begin{eqnarray}
T_{D{K}}^{(0)}&=&\frac{m_K^2}{8\pi^2f_K^4}\left\{3m_K\left(1-\ln\frac{|m_K|}{\lambda}-\ln\frac{m_\eta}{\lambda}\right)\right.\nonumber\\
&&\left.-3\sqrt{m_\eta^2-m_K^2}\arccos\frac{-m_K}{m_\eta}\right.\nonumber\\
&&\left.+\frac{1}{6}g^2\pi
\left(7m_\eta+\frac{6m_\pi^2}{m_\eta+m_\pi}\right) \right\},
\end{eqnarray}
\begin{eqnarray}
T_{D\bar{K}}^{(1)}&=&\frac{m_K^2}{16\pi^2f_K^4}\left\{-m_K\left(3-\ln\frac{m_\pi}{\lambda}
-2\ln\frac{|m_K|}{\lambda}-3\ln\frac{m_\eta}{\lambda}\right)\right.\nonumber\\
&&+\sqrt{m_K^2-m_\pi^2}\ln\frac{m_K+\sqrt{m_K^2-m_\pi^2}}{m_\pi}\nonumber\\
&&-3\sqrt{m_\eta^2-m_K^2}\arccos\frac{m_K}{m_\eta}\nonumber\\
&&\left.+\frac{1}{3}g^2\pi
\left(3m_\eta+\frac{2m_\pi^2}{m_\eta+m_\pi}\right) \right\},
\end{eqnarray}
\begin{eqnarray}
T_{D\bar{K}}^{(0)}&=&\frac{m_K^2}{16\pi^2f_K^4}\left\{3m_K\left(1+\ln\frac{m_\pi}{\lambda}
-2\ln\frac{|m_K|}{\lambda}-\ln\frac{m_\eta}{\lambda}\right)\right.\nonumber\\
&&+3\sqrt{m_K^2-m_\pi^2}\ln\frac{m_K+\sqrt{m_K^2-m_\pi^2}}{m_\pi}\nonumber\\
&&+3\sqrt{m_\eta^2-m_K^2}\arccos\frac{m_K}{m_\eta}\nonumber\\
&&\left.-\frac{1}{3}g^2\pi
\left(5m_\eta+\frac{6m_\pi^2}{m_\eta+m_\pi}\right) \right\},
\end{eqnarray}
\begin{eqnarray}
T_{D_sK}&=&\frac{3m_K^2}{16\pi^2f_K^4}\left\{-m_K\left(1-\ln\frac{m_\pi}{\lambda}
-\ln\frac{m_\eta}{\lambda}\right)\right.\nonumber\\
&&+\sqrt{m_K^2-m_\pi^2}\ln\frac{m_K+\sqrt{m_K^2-m_\pi^2}}{m_\pi}\nonumber\\
&&\left.-\sqrt{m_\eta^2-m_K^2}\arccos\frac{m_K}{m_\eta}+\frac{4}{9}g^2\pi
m_\eta \right\},
\end{eqnarray}
\begin{eqnarray}
T_{D_s\bar{K}}&=&\frac{3m_K^2}{16\pi^2f_K^4}\left\{m_K\left(1-\ln\frac{m_\pi}{\lambda}
-\ln\frac{m_\eta}{\lambda}\right)\right.\nonumber\\
&&+\sqrt{m_K^2-m_\pi^2}\left(i\pi-\ln\frac{m_K+\sqrt{m_K^2-m_\pi^2}}{m_\pi}\right)\nonumber\\
&&\left.-\sqrt{m_\eta^2-m_K^2}\arccos\frac{-m_K}{m_\eta}+\frac{4}{9}g^2\pi
m_\eta \right\},
\end{eqnarray}
\begin{eqnarray}
T_{D\pi}^{(3/2)}&=&\frac{m_\pi^2}{8\pi^2f_\pi^4}\left\{-m_\pi\left(\frac32-2\ln\frac{m_\pi}{\lambda}
-\ln\frac{|m_K|}{\lambda}\right)\right.\nonumber\\
&&-\sqrt{m_K^2-m_\pi^2}\arccos\frac{m_\pi}{m_K}\nonumber\\
&&\left.+\frac{1}{12}g^2\pi (9m_\pi-m_\eta) \right\},
\end{eqnarray}
\begin{eqnarray}
T_{D\pi}^{(1/2)}&=&\frac{m_\pi^2}{8\pi^2f_\pi^4}\left\{m_\pi\left(3-4\ln\frac{m_\pi}{\lambda}
-2\ln\frac{|m_K|}{\lambda}\right)\right.\nonumber\\
&&-\sqrt{m_K^2-m_\pi^2}\left(\frac32\pi-2\arccos\frac{m_\pi}{m_K}\right)\nonumber\\
&&\left.+\frac{1}{12}g^2\pi \left(9m_\pi-m_\eta\right) \right\},
\end{eqnarray}
\begin{eqnarray}
T_{D_s\pi}&=&\frac{m_\pi^2}{24\pi
f_\pi^4}\left\{-3\sqrt{m_K^2-m_\pi^2}-g^2m_\eta \right\},
\end{eqnarray}
\begin{eqnarray}
T_{D\eta}&=&\frac{1}{16\pi
f_\eta^4}\left\{3i\sqrt{m_\eta^2-m_K^2}m_\eta^2-\frac32g^2m_\pi^3\right.\nonumber\\
&&\left.-\frac16g^2(4m_\eta^2-m_\pi^2)m_\eta+2g^2m_K^3 \right\},
\end{eqnarray}
\begin{eqnarray}
T_{D_s\eta}&=&\frac{1}{24\pi
f_\eta^4}\left\{9i\sqrt{m_\eta^2-m_K^2}m_\eta^2\right.\nonumber\\
&&\left.-g^2(4m_\eta^2-m_\pi^2)m_\eta+6g^2m_K^3 \right\}.
\end{eqnarray}

In deriving the above analytical expressions, we have used the
dimensional regularization and minimal subtraction scheme. Here we
retain the chiral logarithm terms. $\lambda\sim 1$ GeV denotes the
scale of chiral symmetry breaking. In these T-matrices, the
corrections proportional to $g^2$ come from intermediate vector
meson contributions (the last six diagrams depicted in Fig
\ref{figg}) and the divergences from these diagrams cancel out.
The fourth diagram generates imaginary parts for $T_{DK}^{(1)}$,
$T_{D_s\bar{K}}$, $T_{D\eta}$ and $T_{D_s\eta}$.

From ${\cal L}^{(3)}_{H\phi}$, we obtain the counter terms:
\begin{eqnarray}
&T_{D{K}}^{(1)}=0,\quad T_{D{K}}^{(0)}=16\kappa^r\frac{
m_K^3}{f_K^2},&
\nonumber \\
& T_{D\bar{K}}^{(1)}=-8\kappa^r\frac{ m_K^3}{f_K^2},\quad
T_{D\bar{K}}^{(0)}=8\kappa^r\frac{
m_K^3}{f_K^2},&\nonumber\\
&T_{D_sK}=-8\kappa^r\frac{ m_K^3}{f_K^2},\quad
T_{D_s\bar{K}}=8\kappa^r\frac{
m_K^3}{f_K^2},&\nonumber\\
&T_{D\pi}^{(3/2)}=-8\kappa^r\frac{ m_\pi^3}{f_\pi^2},\quad
T_{D\pi}^{(1/2)}=16\kappa^r\frac{ m_\pi^3}{f_\pi^2},&\nonumber\\
&T_{D_s\pi}=T_{D\eta}=T_{D_s\eta}=0,&
\end{eqnarray}
where $\kappa^r=\kappa^r(\lambda)=\kappa-\frac34\frac{L}{f^2}$ is
the renormalized LEC with
$L=\frac{\lambda^{d-4}}{16\pi^2}[\frac{1}{d-4}+\frac12(\gamma_E-1-\ln{4\pi})]$.

One may verify that the above threshold T-matrices satisfy the
crossing symmetry:
\begin{eqnarray}
T_{D\bar{K}}^{(1)}&=&\frac12[T_{DK}^{(1)}+T_{DK}^{(0)}]_{m_K\rightarrow-m_K},\\
T_{D\bar{K}}^{(0)}&=&\frac12[3T_{DK}^{(1)}-T_{DK}^{(0)}]_{m_K\rightarrow-m_K},\\
T_{D_s\bar{K}}&=&[T_{D_sK}]_{m_K\rightarrow-m_K},
\end{eqnarray}
and the SU(3) relations ($m_u=m_d=m_s$)
\begin{eqnarray}
&T_{DK}^{(1)}=T_{D_s\pi}=\frac12[T_{D\bar{K}}^{(1)}+T_{D\bar{K}}^{(0)}],&\\
&T_{D_sK}=T_{D\pi}^{(3/2)}=T_{D\bar{K}}^{(1)},&\\
&T_{D_s\bar{K}}=\frac13[T_{D\pi}^{(3/2)}+2T_{D\pi}^{(1/2)}]=\frac12[T_{DK}^{(1)}+T_{DK}^{(0)}],&\\
&T_{D_s\pi}+T_{D_s\eta}=\frac13[2T_{D\pi}^{(3/2)}+T_{D\pi}^{(1/2)}]+T_{D\eta}.&
\end{eqnarray}
These relations are very useful to cross-check the results.

\section{Numerical results}\label{sec4}

There are four parameters to be determined in our T-matrices: $c_1$,
$C_1$, $C_0$ and $\kappa^r$. We extract $c_1$ from the mass
splitting between heavy mesons within the same doublet:
\begin{eqnarray}
4c_1\stackrel{\circ}{M}(m_K^2-m_\pi^2)=\frac12(M_{D_s}^2-M_D^2+M_{D_s^*}^2-M_{D^*}^2),\nonumber\\
\end{eqnarray}
where $\stackrel{\circ}{M}$ is the heavy meson mass in the heavy
quark limit and chiral limit. We take the value
$\stackrel{\circ}{M}\approx 1918$ MeV for the charmed system
\cite{HLutz}. Using $M_D=1867.2$ MeV, $M_{D_s}=1968.5$ MeV,
$M_{D^*}=2008.6$ MeV, and $M_{D_s^*}=2112.3$ MeV \cite{PDG}, we
get $c_1=0.47$ GeV$^{-1}$.

Up to now, there does not exist any available experimental
measurement of these scattering lengths. A recent lattice
simulation yielded $a_{D\pi}^{(3/2)}=-0.16(4)$ fm,
$a_{D\bar{K}}^{(1)}=-0.23(4)$ fm, $a_{D_s\pi}=0.00(1)$ fm and
$a_{D_sK}=-0.31(2)$ fm \cite{latDK}. In order to determine the
other three LECs, we use these values as inputs.

From $a_{D\pi}^{(3/2)}$, $a_{D_sK}$ and $a_{D_s\pi}$, we get
$C_1=-0.63$ GeV$^{-1}$, $C_0=5.64$ GeV$^{-1}$ and $\kappa^r=-0.33$
GeV$^{-2}$. We have used $m_\pi=139.6$ MeV, $m_K=493.7$ MeV,
$m_\eta=547.8$ MeV, $f_\pi=92.4$ MeV, $f_K=113$ MeV,
$f_\eta=1.2f_\pi$ and $\lambda=4\pi f_\pi$. The coupling constant
$g=0.59\pm0.07\pm0.01$ was determined from the width of $D^{*+}$
\cite{widthDstar,gcoup}.

We present our numerical estimations for T-matrices and scattering
lengths in Table \ref{results-c}. For comparison, we also list
several scattering lengths appearing in the literature.

\begin{table*}
\begin{tabular}{c|cccccc}
\hline\hline
                      &${\cal O}(p)$&${\cal O}(p^2)$&${\cal O}(p^3)$&Total&Scattering lengths& Other results\\\hline
$T_{DK}^{(1)}$        & 0           & 9.4           & $-1.5+5.6i$   &$7.9+5.6i$    & $0.25+0.17i$ \\
$T_{DK}^{(0)}$        &15.3         & $-2.4$        & $0.1$        &13.0        & 0.41     & $0.98\pm0.20$ \cite{beve} \\
$T_{D\bar{K}}^{(1)}$& $-7.6$   &  $-2.4$       & $-1.2$         & $-11.2$          &$-0.35$    \\
$T_{D\bar{K}}^{(0)}$& $7.6$    &  $21.2$       & $2.5$         & $31.4$          &$0.99$   \\
$T_{D_s{K}}$& $-7.6$  &  $-2.4$       & $0.3$         & $-9.7$          &$-0.31$ (input)  \\
$T_{D_s\bar{K}}$ & $7.6$   &  $-2.4$       & $-1.5+8.3i$         & $3.8+8.3i$          &$0.12+0.27i$  \\
$T_{D\pi}^{(3/2)}$     & $-3.2$     & $-0.3$        & $-0.8$         &$-4.3$      & $-0.16$ (input) \\
$T_{D\pi}^{(1/2)}$     & $6.5$     & $-0.3$        & $0.3$         &$6.4$      & $0.24$ & $0.41\pm0.06$ \cite{nieves}\\
$T_{D_s\pi}$          & 0.0        & $1.1 $        & $-1.1$         &$0.0$      & 0.00 (input) \\
$T_{D\eta}$           & 0.0        & $9.9 $        & $1.2+5.5i$     &$11.0+5.5i$      & $0.34+0.17i$ \\
$T_{D_s\eta}$         & 0.0        & $-13.8 $        & $0.5+11.1i$     &$-13.3+11.1i$      & $-0.41+0.35i$ \\
\hline\hline
\end{tabular}
\caption{Threshold T-matrices for the elastic scattering of pions
and charmed pseudoscalar mesons in unit of fm with the scale
$\lambda=4\pi f_\pi$. For the results from a unitarized method, one
may consult Ref. \cite{guo}.}\label{results-c}
\end{table*}

The heavy quark symmetry works better for bottom systems. We would
like to estimate the S-wave scattering lengths for the elastic
pion-bottom meson interactions as well.

In the bottom case, one extracts $c_1=0.39$ GeV$^{-1}$ with the mass
\cite{PDG} $M_B=5279.3$ MeV, $M_{B_s}=5366.3$ MeV, $M_{B^*}=5325.1$
MeV, $M_{B_S^*}=5412.8$ MeV and
$\stackrel{\circ}{M}=\frac12(M_B+M_{B_s})=5322.8$ MeV. The heavy
quark flavor symmetry tells us that we may still use the above
determined values for $C_1$, $C_0$ and $\kappa^r$. The coupling
constant $g$ from a recent unquenched lattice result is
$0.516(5)_{stat}(33)_{chiral}(28)_{pert}(28)_{disc}$ \cite{latBpi}.
We use $g=0.52$ to estimate the values. The calculated scattering
lengths are presented in Table \ref{results-b}.

\begin{table*}
\begin{tabular}{c|cccccc}
\hline\hline
                      &${\cal O}(p)$&${\cal O}(p^2)$&${\cal O}(p^3)$&Total&Scattering lengths & Other results\\\hline
$T_{\bar{B}K}^{(1)}$        & 0           & 9.4           & $-1.4+5.6i$   &$8.0+5.6i$    & $0.29+0.20i$ \\
$T_{\bar{B}K}^{(0)}$        &15.3         & $-2.4$        & $-0.5$        &12.3        & 0.45      \\
$T_{\bar{B}\bar{K}}^{(1)}$& $-7.6$   &  $-2.4$       & $-1.4$         & $-11.4$          &$-0.42$    \\
$T_{\bar{B}\bar{K}}^{(0)}$& $7.6$    &  $21.2$       & $3.0$         & $31.9$          &$1.16$   \\
$T_{\bar{B}_s{K}}$& $-7.6$  &  $-2.4$       & $-0.1$         & $-10.1$          &$-0.37$   \\
$T_{\bar{B}_s\bar{K}}$ & $7.6$   &  $-2.4$       & $-1.8+8.3i$         & $3.4+8.3i$          &$0.12+0.30i$  \\
$T_{\bar{B}\pi}^{(3/2)}$     & $-3.2$     & $-0.3$        & $-0.8$         &$-4.3$      & $-0.17$  \\
$T_{\bar{B}\pi}^{(1/2)}$     & $6.5$     & $-0.3$        & $0.2$         &$6.4$      & $0.25$ & $0.37\pm0.37$ \cite{nieves}\\
$T_{\bar{B}_s\pi}$          & 0        & $1.1 $        & $-1.1$         &$0.03$      & 0.00  \\
$T_{\bar{B}\eta}$           & 0        & $9.4 $        & $0.9+5.5i$     &$10.3+5.5i$      & $0.37+0.20i$ \\
$T_{\bar{B}_s\eta}$         & 0        & $-12.8 $        & $0.3+11.1i$     &$-12.4+11.1i$      & $-0.45+0.40i$ \\
\hline\hline
\end{tabular}
\caption{Threshold T-matrices for the elastic scattering of pions
and bottom pseudoscalar mesons in unit of fm with the scale
$\lambda=4\pi f_\pi$. Here $\bar{B}=(\bar{B}^0,
B^-)^T$.}\label{results-b}
\end{table*}

In principle, the scale dependence from the chiral loop is
compensated by LECs. By varying the scale $\lambda=4\pi f_\pi$ to
$\lambda=m_\rho=775.49$ MeV, one may check the analytic independence
on it. By repeating the above procedure, we get $C_1=-0.75$
GeV$^{-1}$, $C_0=5.76$ GeV$^{-1}$ and $\kappa^r=-0.21$ GeV$^{-2}$ at
this lower scale. We show the numerical estimations for the charmed
systems in Table \ref{results-scale}. The scale independence is
illustrated if one compares the results with those in Table
\ref{results-c}.

\begin{table*}
\begin{tabular}{c|ccccc}
\hline\hline
                      &${\cal O}(p)$&${\cal O}(p^2)$&${\cal O}(p^3)$&Total&Scattering lengths\\\hline
$T_{DK}^{(1)}$        & 0           & 9.4           & $-1.5+5.6i$   &$7.9+5.6i$    & $0.25+0.17i$ \\
$T_{DK}^{(0)}$        &15.3         & $-2.8$        & $-0.8$        &11.7        & 0.37      \\
$T_{D\bar{K}}^{(1)}$& $-7.6$   &  $-2.8$       & $-0.7$         & $-11.2$          &$-0.35$    \\
$T_{D\bar{K}}^{(0)}$& $7.6$    &  $21.7$       & $2.1$         & $31.4$          &$0.99$   \\
$T_{D_s{K}}$& $-7.6$  &  $-2.8$       & $0.7$         & $-9.7$          &$-0.31$ (input)  \\
$T_{D_s\bar{K}}$ & $7.6$   &  $-2.8$       & $-1.9+8.3i$         & $2.9+8.3i$          &$0.09+0.27i$  \\
$T_{D\pi}^{(3/2)}$     & $-3.2$     & $-0.3$        & $-0.8$         &$-4.3$      & $-0.16$ (input) \\
$T_{D\pi}^{(1/2)}$     & $6.5$     & $-0.3$        & $0.2$         &$6.3$      & $0.23$ \\
$T_{D_s\pi}$          & 0        & $1.1 $        & $-1.1$         &$0.0$      & 0.00 (input) \\
$T_{D\eta}$           & 0        & $9.7 $        & $1.2+5.5i$     &$10.8+5.5i$      & $0.33+0.17i$ \\
$T_{D_s\eta}$         & 0        & $-14.5 $        & $0.5+11.1i$     &$-14.1+11.1i$      & $-0.44+0.35i$ \\
\hline\hline
\end{tabular}
\caption{Threshold T-matrices for the elastic scattering of pions
and charmed pseudoscalar mesons in unit of fm with a lower scale
$\lambda=m_\rho$.}\label{results-scale}
\end{table*}

\section{Discussions}\label{sec5}

We discuss the results at the scale $\lambda=4\pi f_\pi$. By
comparing the values in Table \ref{results-b} and \ref{results-c},
one notices the difference between the results of pions-bottom
mesons and those of pions-charmed mesons is small. Larger difference
will appear after including the recoil corrections. We focus mainly
on charmed systems in the following discussions.

The positive sign of $a_{DK}^{(1)}$, $a_{DK}^{(0)}$,
$a_{D\bar{K}}^{(0)}$, $a_{D\pi}^{(1/2)}$ and $a_{D\eta}$ indicates
that the interactions for these channels are all attractive. For
the isoscalar $D-K$ channel, the attraction is relatively strong.
However, further exploration of the phase shifts of the elastic
$DK$ scattering is required in order to answer whether the $DK$
interaction is strong enough to form a bound state such as a $DK$
molecular state.

The result for the isoscalar $D-\bar{K}$ channel is very
interesting. Since no quark pair annihilation occurs in this
channel, the relatively large scattering length implies that the
possibility to form a four-quark resonance state is not excluded.
If the strong attraction is confirmed in future studies, this
channel is certainly worthwhile further exploration.

The $D-\eta$ channel is also interesting. Twenty years ago, the
formation of $\eta$-mesic nucleus \cite{etamesic} was proposed based
on the observation that $\eta$ is neutral and the strong force
between $\eta$ and $N$ is attractive. Later the possibility of the
$\eta$-hypernuclei was also discussed \cite{etahyper}. In Ref.
\cite{Dmesic}, the bound state of a $D$ meson and $^{208}P_b$ was
studied. Now, with the observation that both $D\eta$ and $\eta N$
are attractive, one may guess the formation of a $\eta-D-$heavy
nucleus state is also possible. However, such a bound state would be
very difficult to be detected experimentally \cite{qmc}.

Contrary to the $D-\eta$ channel, the scattering length
$a_{D_s\eta}$ is negative, which indicates the interaction is
repulsive. In order to get a positive $a_{D_s\eta}$ and nearly
vanishing $a_{D_s\pi}\sim 0$, one requires $C_1>1.4$ GeV$^{-1}$
and $C_0<3.6$ GeV$^{-1}$.

From Table \ref{results-c}, we note that the chiral expansion
converges well in four channels: isoscalar $DK$, isovector
$D\bar{K}$, $D_sK$ and $D_s\bar{K}$. The convergence of
$T_{DK}^{(1)}$, $T_{D\eta}$ and $T_{D_s\eta}$ will be manifest
when the ${\cal O}(p^4)$ corrections are included. Although the
isoscalar $D-\bar{K}$ channel is very interesting, there is a
critical problem of convergence. The contribution from the second
order is very large. Future accurate determination of LECs is
probably helpful to diminish this contribution.

Future improvements may be made for the present calculation. We
have used the recent lattice results in full QCD to estimate the
LECs where the lattice pion mass is large \cite{latDK}. Due to the
lack of data, the deduced LECs have large uncertainties. Since no
$J^P=0^+$ resonance close to the $D\pi$ threshold has been
reported, this channel is ideal to determine the LECs. To test
HM$\chi$PT in the scattering problem, the experimental
determination of pion-heavy meson scattering lengths is strongly
expected. On the other hand, further lattice simulations can
provide more reliable inputs. We hope our chiral corrections may
be useful when performing chiral extrapolations.

One may also improve the analytical expressions by including $1/M$
corrections and extending the calculation to higher orders in
future exploration. In addition, the effects of the nearby $0^+$
resonances may affect the predictions. One can consider them using
the non-perturbative methods.

Finally, we mention that the scattering lengths of pions with
heavy anti-mesons. They are easy to get through the C parity
transformation
\begin{eqnarray}
T_{\bar{H}K}^{(I)}=T_{H\bar{K}}^{(I)},\quad
T_{\bar{H}\bar{K}}^{(I)}=T_{HK}^{(I)},\quad
T_{\bar{H}{\pi/\eta}}^{(I)}=T_{H{\pi/\eta}}^{(I)},
\end{eqnarray}
where $I$ is the total isospin and $H$ ($\bar{H}$) denotes the heavy
meson (anti-meson).

\section*{Acknowledgments}

We are grateful to Professor Zong-Ye Zhang and Professor Chuan Liu
for helpful discussions. This project was supported by the National
Natural Science Foundation of China under Grants 10775146, 10805048,
10705001, 10625521 and 10721063, the China Postdoctoral Science
foundation (20070420526), K.C. Wong Education Foundation, Hong Kong
and in part by the \emph{Funda\c{c}\~{a}o para a Ci\^{e}ncia e a
Tecnologia} \/of the \emph{Minist\'{e}rio da Ci\^{e}ncia, Tecnologia
e Ensino Superior} \/of Portugal, under contract
SFRH/BPD/34819/2007.

\end{document}